\documentclass{optica-article}

\journal{opticajournal} 

\articletype{Research Article}

\usepackage[utf8]{inputenc}
\usepackage[T1]{fontenc}

\usepackage{floatrow}
\usepackage{amsmath}
\usepackage{graphicx}
\usepackage{parskip}

\usepackage[labelfont=bf]{caption}

\begin{document}

\title{Broadband Continuous Spectral Control of a Single Wavelength Polymer-Based Solid-State Random Laser}

\author{Bhupesh Kumar,\authormark{1,2} Sebastian A. Schulz,\authormark{2} and Patrick Sebbah\authormark{1,*}}

\address{\authormark{1}Department of Physics, The Jack and Pearl Resnick Institute for Advanced Technology, Bar-Ilan University, Ramat-Gan, 5290002 Israel\\
\authormark{2}SUPA, School of Physics and Astronomy, University of St. Andrews, Fife, KY16 9SS, UK\\}
\email{\authormark{*}patrick.sebbah@biu.ac.il} 

\begin{abstract*} 
We demonstrate temperature-controlled spectral tunability of a partially-pumped single-wavelength random laser in a solid-state random laser based on DCM (4-dicyanomethylene-2-methyl-6-(p-dimethylaminostyryl)-4H-pyran) doped PMMA (polymethyl methacrylate) dye. By carefully shaping the spatial profile of the pump, we first achieve low-threshold, single-mode random lasing with excellent side lobes rejection. Notably, we show how temperature-induced changes in the refractive index of the PMMA-DCM layer result in a blue-shift of this single lasing mode. Continuous tunability of the lasing wavelength is demonstrated over a 8nm-wide bandwidth.

\end{abstract*}

\section{Introduction}
Random lasers are unconventional laser sources in which feedback is provided by randomly-distributed scattering particles. In the past two decades, random lasers have been the subject of intense theoretical and experimental study \cite{Cao99, Vanneste01, Matos07, Sapienza19, Lodhal14, rotter16, gollner15, yzhang13, abaie17}. Tunability and directionality are important features that determine the application scope of any laser device in fields such as integrated spectroscopy, remote sensing, and optical communication\cite{wiersma08}. Single wavelength random laser tunability is challenging due to the random and multimode nature of the emission spectrum,  fluctuations in the emission spectrum, and lack of precise, non-invasive and reversible tuning mechanism. 
Spectral tunability in random lasers has been demonstrated via multiple mechanisms, including   optical fiber-based random lasers\cite{zhijia17}, pump size control or scatterer concentration variation \cite{bkumar22}, gain medium thickness variation \cite{tong19}, dye molecule selection \cite{ziegler16, wang17}, or mechanical stretching \cite{zhai15, Lee18}. Other tuning mechanisms include engineering absorption of light emission\cite{RGSEI11}, switching modes associated with different lengths of silver nanorods in plasmonic random lasers\cite{junxi22}, as well as electric-field-induced tunability \cite{qinghai09}.
However, all these mechanisms have been limited to the spectral tuning of broad linewidth or multimode random lasers. Tunable single-mode random laser has been reported in rare-earth-doped fiber random laser, emitting exclusively in the mid-near-infrared \cite{ruo22,jing21,han21, du15}. Single-mode random lasing temperature-based tunability in the visible has been demonstrated in liquid crystal-embedded random lasers, but tunability was found to be limited to a few nanometers by the nematic-isotropic transition temperature \cite{song07}. 
To the best of our knowledge, continuous broadband tuning of a singlemode random laser in the visible as not yet been reported.
\\
 Recently, we have demonstrated that using an iterative pump shaping optimization method, selective excitation of a particular mode of the multimode emission spectrum and single-mode operation with high side-lobe rejection can be achieved \cite{patrick14, NicolasPRL,bhupesh21}. However, it is not possible to achieve lasing at any arbitrary desired wavelength, but only at wavelengths corresponding to lasing modes of the discrete multimode emission spectrum. Overcoming this limitation would add tunability to this technique and offer full spectral control of the random laser.
In this paper, we report single-wavelength continuous broadband spectral tuning in a dye-doped solid-state random laser. Random lasers (RLs) typically produce multimode coherent light due to the random distribution of scattering particles. By combining the technique of spatial pump shaping together with the negative thermal coefficient of the refractive index of PMMA polymer, we are able to achieve single wavelength spectral control of a random laser on a single device. Specifically, we show continuous broadband tunability over a bandwidth of 8 nm.

\section{Sample fabrication}
Disordered structures were fabricated using e-beam lithography on a 600 nm layer of poly(methyl methacrylate) or PMMA that was doped with 5$\%$ weight of DCM laser dye. The fluorescence spectrum of the dye is centered around 600 nm; it has a high fluorescence quantum yield and a large stoke shift (100 nm), which means it does not reabsorb the emitted light. The PMMA used had a molecular weight of 49500 and was used at a concentration of 6$\%$ weight in anisole.

\begin{figure}[htbp]   
  \centering \includegraphics[width=8cm]{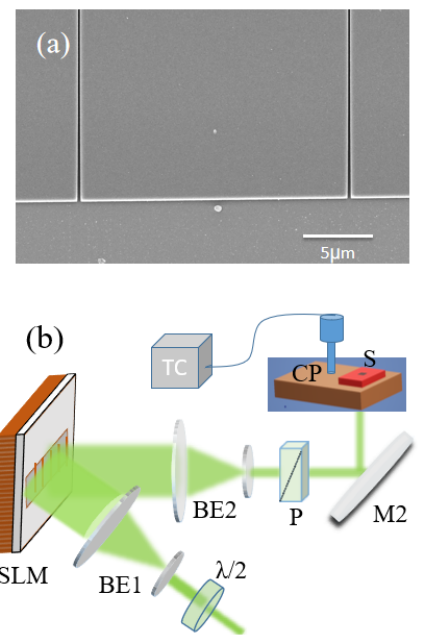}
\caption{(a) High resolution SEM image of the small section of the sample from the top showing air grooves carved into the PMMA-DCM layer. (b) Schematic of the experimental setup.  BE1 and BE2:  Beam expanders; P: Polarizer, SLM: Spatial light modulator; M2:  Mirror; S: Sample; TC: Temperature controller; CP: Copper plate; }
\end{figure}

The active layer was obtained by spin-coating a doped-polymer solution at 1000 rpm for 60 seconds on a fused silica wafer (Edmund optics) and post-baking it at 120°C for 2 hours in an oven\cite{bhupesh21}.
The silica wafer has a refractive index of 1.45, compared to 1.54 for the PMMA-DCM layer. To fabricate one dimensional(1D)-disorder samples, 125 randomly distributed parallel grooves, each 200 nm wide and 50 µm long, covering a total length of 1000 $\mu$m,  were carved into the PMMA-DCM layer using e-beam lithography (CRESTEC/CABL-9000C) (see Fig. 1(a)), resulting in dielectric sections with an average length of 8 µm separated by air gaps of 200 nm. The fabrication method ensures a high refractive index contrast of 0.54 between air grooves and the polymer layer, which helps in achieving random lasing action at a low threshold. 
We require manuscripts to identify a single corresponding author. 

\section{Experimental Setup}
A schematic of the experimental setup is shown in Fig.1(b). The setup consists of a frequency-doubled mode-locked Nd:YAG laser (EXPLA PL2230: 532 nm, 20 ps, maximum output energy 28 mJ, repetition rate 1-50 Hz). The beam is expanded 5X and spatially modulated by a 1952 x 1088 pixel reflective spatial light modulator (SLM) (Holoeye HES 6001, pixel size 8.0 µm). The SLM is placed in the object plane of a telescope with 4X reduction and is imaged on the sample. This setup is used to create a laser strip whose width and length are adjusted according to the sample dimensions. The disorder structure is precisely aligned with the laser strip under a fixed stage Zeiss microscope (AxioExaminer A1) and imaged using an Andor Zyla sCMOS camera (22 mm diagonal view, 6.5 µm pixel size) using a 10X objective. The laser emission is collected through a multimode optical fiber connected to a Horiba iHR550 imaging spectrometer equipped with a 2400 $mm^{-1}$ grating and Synapse CCD detection system (sampling rate 1 MHz, 1024 x 256 pixels, 26 µm pixel pitch). The entrance slit is 50 µm, resulting in a spectral resolution of 20 pm.

\begin{figure}[htbp]   
\centering \includegraphics[width=10cm]{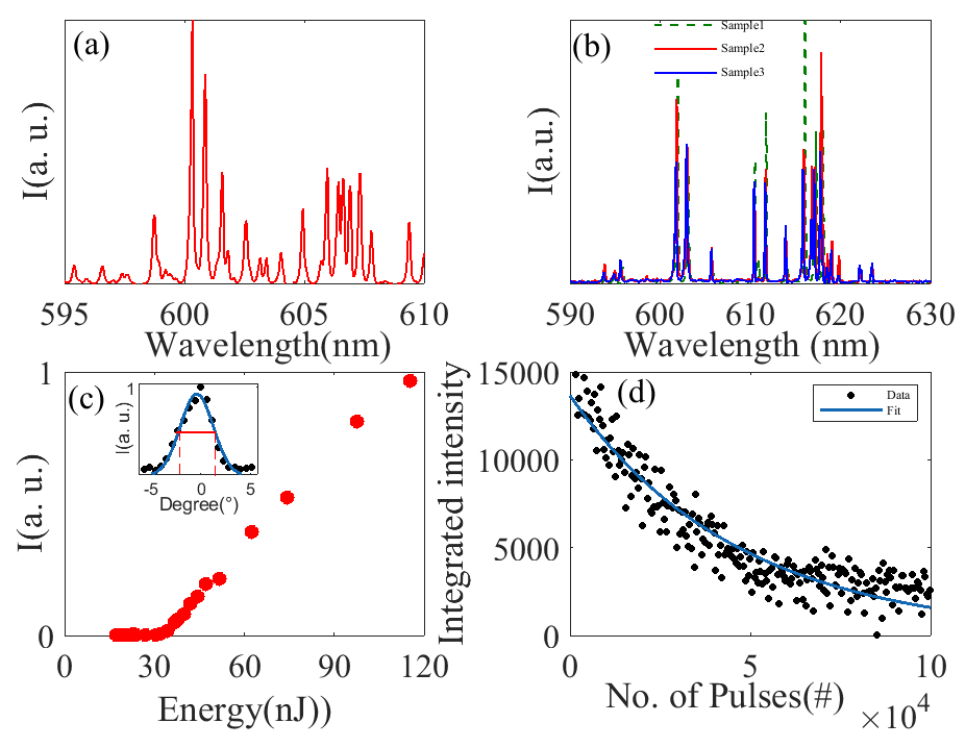}
\caption{ (a) Multimode Random laser emission spectra obtained after pumping the sample uniformly above the threshold, sample length 1000 $\mu$m, pump size 1000 $\mu$m X 50 $\mu$m .  (b) Emission spectra were recorded for 3 different samples of identical disorder configuration at same pump power. (c) The integrated intensity of the emission spectrum with increasing Pump energy (nJ). inset: Angular distribution of the integrated output intensity of random laser emission at a distance of 10 cm from the sample edge, the blue line is a Gaussian fit. (d) Integrated intensity as a function of the number of pump pulses for the random laser. The pump energy is about 50 nJ for a uniform pump of size 450 $\mu$m X 50 $\mu$m at a repetition rate of 10 Hz. The blue line is an exponential fit.}
\end{figure}

 \begin{figure}[htbp]   
  \centering \includegraphics[width=8cm]{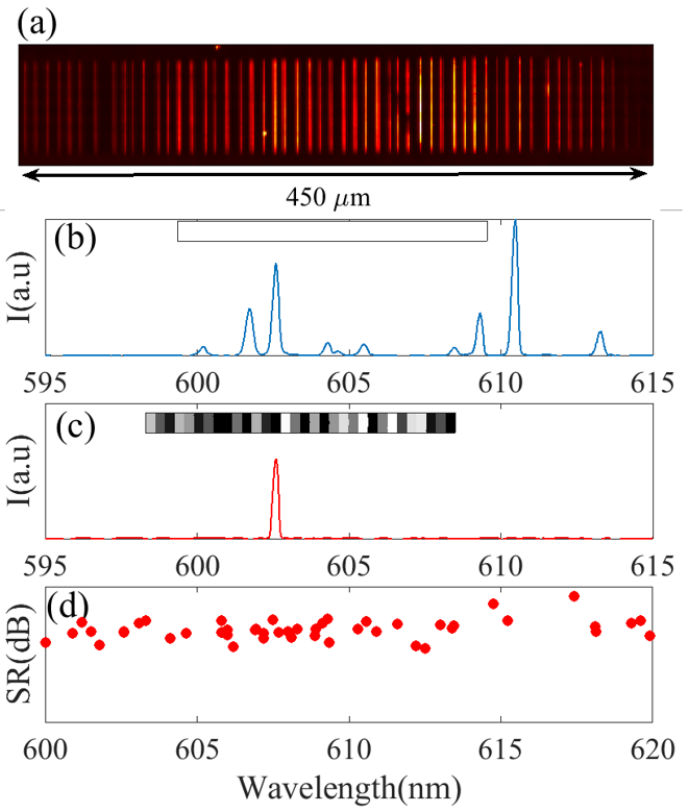}
\caption{(a) Optical microscope image of field intensity distribution near the sample surface when pumped above the threshold. (b) Multimode emission spectra for a uniform pump strip of size 450 $\mu$m X 
50 $\mu$m. (c) Singlemode lasing at $\lambda$= 602.60 nm after an iterative pump optimization process. Inset shows the optimized pump profile in grayscale. 
(d) Iterative optimization of pump shaping has been
applied to select 50 lasing modes. The sidelobe rejection ratio(SR) for each individually selected lasing mode w.r.t maximum noise count in the emission spectrum is plotted in log representation (dB).}
\end{figure}

\section{Results}
When the sample undergoes optical pumping, amplified spontaneous emission experiences multiple scattering. This provides in turn coherent optical feedback, enabling random lasing oscillations with sharp emission linewidth. Once the pump intensity reaches above the lasing threshold, the output intensity increases manifold. This nonlinear increase in output intensity confirms the onset of random lasing action.  By linear fit, the lasing threshold value is found to be 35 nJ for a pump size of 450 $\mu$m X 50 $\mu$m (Fig.2(c)). Above the threshold, multimode lasing is achieved (illustrated in Fig.2(a)). The resulting spectrum displays randomly positioned, distinct lasing peaks, each exhibiting a typical linewidth of 0.2 nm, constrained by the resolution of the spectrometer.
\begin{figure}[htbp]   
 \centering \includegraphics[width=10cm]{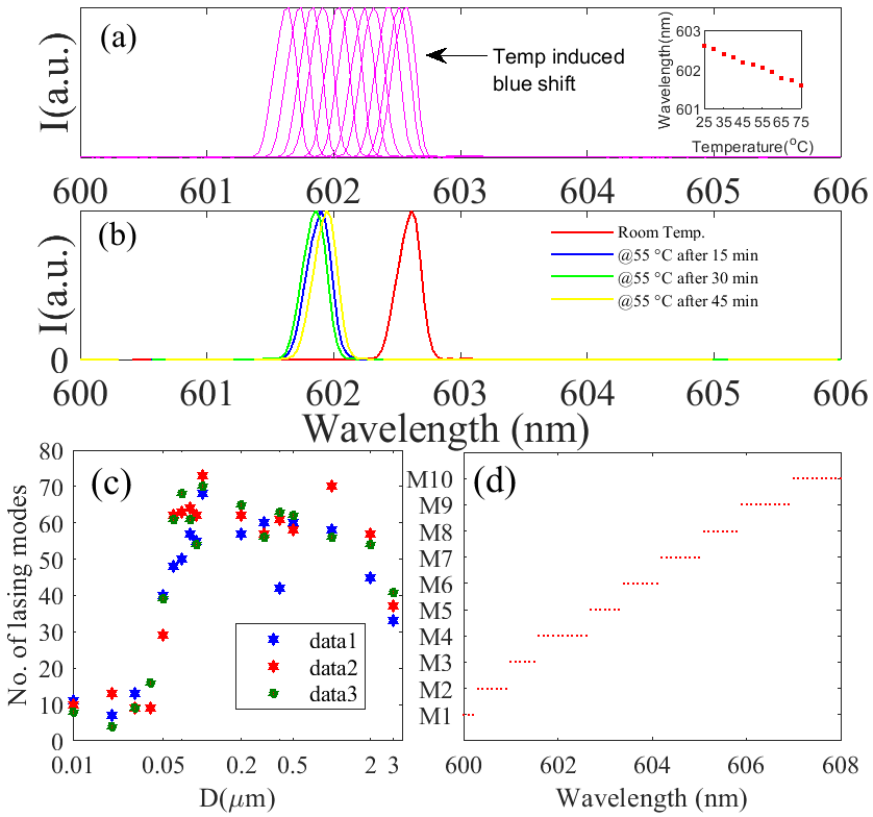}
\caption{
(a) The spectral blue shift of 1 nm in the single mode lasing spectrum at $\lambda$ = 602.60 nm with increasing temperature of the polymer layer. The figure in the inset shows the temperature vs wavelength plot.(b) Emission spectrum stability at constant temperature. The emission spectrum was recorded at an interval of 15, 30, 45 minutes when the sample was heated to a constant temperature of 55\textdegree C. (c) Number of lasing modes for three different sample sets plotted as a function of increasing deviation (D) from the mean periodic position of 8 $\mu m$ in a sample of length 1000 $\mu m$ having 125 air grooves. (d) Continuous tunability over a bandwidth of 8 nm by  tuning of 10 different individually selected lasing modes M1= 600.2 nm, M2= 600.9 nm, M3= 601.50 nm, M4= 602.6 nm, M5= 603.3 nm, M6= 604.1 nm, M7= 605.00 nm, M8= 605.800 nm, M9= 606.900 nm, M10= 608.00 nm.}
\end{figure}
The directivity of our random laser is assessed, with half of the emission intensity confined within $\pm$ 2.5 degrees from the center of the sample. This is found by scanning a 20X microscope objective placed at a distance of 10 cm from the sample edge to collect emission in a direction orthogonal to the sample length. Directional measurements are shown in inset Fig.2(c).\\ 
We demonstrate spectral reproducibility by manufacturing three samples with identical disorder configurations on the same substrate. These samples are then subjected to uniform pumping at the same pump power, and their respective emission spectra are recorded. Remarkably, all three samples yield identical emission spectra, as depicted in Fig. 2(b). The high degree of spectral reproducibility is attributed to the highly precise e-beam nanolithography fabrication technique. \\
We also examined the photostability of our quasi-1D random laser by subjecting the sample to uniform pumping using the same experimental conditions as described earlier. The pump energy was held constant at 50 nJ, employing a pump size of 450 µm x 50 µm, and a pump laser repetition rate of 10 Hz. In Figure 2(c), we present the integrated emission intensity graphed against the laser pulse number. The photostability of the laser was observed to be 28380 pulses, equivalent to approximately 47.3 minutes, during which the integrated laser emission intensity reduced to half of its original integrated lasing intensity.\\
To achieve singlemode lasing at any of the lasing wavelengths of the multimode emission spectrum(Fig.3(b)), we apply a nonuniform intensity profile to the pump beam by modulating the SLM. We implement an iterative optimization method \cite{patrick14,bhupesh21} to find the optimal pump profile for which modes other then the target mode are suppressed. Here we use Nelder-Mead simplex algorithm implemented in the fminsearch function of MATLAB to optimize the pump profile. We slightly modified the fminsearch function by setting the initial step (usually the $\delta$ parameter in fminsearch) to 1.0 in order to explore a larger region of the 32-dimensional space. The number of pixels (32) was chosen as a compromise between sensitivity and computation time. The Matlab-generated image that is sent to the SLM is made up of 32 intensity blocks, each of which is encoded on 256 levels of grey and projected onto the sample. To start the optimization process, we choose 32 column-vectors $V_{i}$ from the 32 x 32 binary Hadamard matrix as the initial vertices.

The pump profile P(x) is therefore written as
\begin{center}
$ \Huge{ P(x)}= \Huge{\frac{1}{255} \sum_{i=1}^{32} \beta_{i}V_{i}}$
\end{center}
where $\beta_{i}$ takes discrete values in the range [0,255]. Each vector $\beta_{i}$ corresponds to a particular pump profile associated with a particular emission spectrum $I\left(\lambda\right)$. 
The optimization algorithm uses the inverse of the extinction ratio, $\eta = I_{t}/I_{o}$, as its cost function and aims for its minimization, where $I_{t}$ is the peak intensity of the targeted mode and $I_{o}$ is the highest intensity among all other modes except the targeted mode. To find the optimized pump profile, the algorithm iteratively generates a new pump profile, applies it to the pump stripe, acquires the emission spectrum averaged over 10 shots, and computes the cost function. Mode selection and single mode operation at the lasing wavelength of $\lambda$ = 602.60 nm are achieved after the convergence of the iterative optimization, as shown in Figure 3(c). After 250 iterations, the algorithm converges to a pump profile that suppresses modes other than the targeted mode. The selected lasing mode has a sidelobe rejection ratio (measured as the ratio of the peak intensity count of the target mode w.r.t. maximum noise level in the spectrum) of 800, which corresponds to 53 dB. The next highest intensity count, except for the target mode, in the emission spectrum is 4 counts, which is close to the average noise level present in the spectrum. We also tested the robustness of our pump profile optimization method to achieve single mode lasing by selecting 50 different lasing modes on different samples.  Figure 3(d) shows the sidelobe rejection of selected modes. All the modes are selected with an excellent side-lobe rejection of more than 40 dB.  
\\
Next, we demonstrate how the emission wavelength of the selected lasing mode can be tuned by changing the temperature of the sample. RL sample was positioned on a copper plate featuring a rectangular hole. Placement of the sample over the hole, enabled us to pump the RL sample from the bottom. The copper plate was connected to a heating probe to heat the sample and a sensor to monitor the temperature.
Figure 4(a) shows the blue shift of single-mode random laser emission initially centered at $\lambda$ = 602.60 nm, when temperature is increased. The inset in figure 4(a) shows the linear temperature dependence of the spectral shift. A linear fit gives a slope of d$\lambda$/dT= -0.02 nm/\textdegree C., which means that a 5 \textdegree C increase results in a spectral shift of 0.1 nm. Overall, we obtained a total shift of 1 nm between 25 \textdegree C to 75 \textdegree C.
When the sample is cooled down to 25 \textdegree C, the initial spectral position is recovered and the process is perfectly reversible. This behavior is easily explained by the fact that PMMA polymer has a negative thermal coefficient of refractive index\cite{xaio01, duarte03, duarte02}. Its refractive index therefore decreases with increasing temperature, which results in a decrease in the optical path length within the polymer layer and a blue-shift. This process is reversible. \\
We also performed a spectral stability test at a constant temperature higher than room temperature in a minimum airflow lab environment. An optimally-pumped lasing sample
emitting light at $\lambda$= 602.60 nm is supplied with constant heat to achieve a wavelength shift of 0.60 nm at sample temperature of 55 \textdegree C.
The emission spectra are recorded at 3 intervals of 15 minutes each.
Lasing emission at @$\lambda$= 602.00 nm remains stable within $\pm$ 0.1 nm as shown in Fig.4(b).
\\
Since all lasing modes within the multimode spectrum can be selected individually \cite{bhupesh21}, the tuning range can be in principle extended over the whole gain curve by hopping from mode to mode. To demonstrate continuous tunability by mode hopping over a broader frequency range, we need first to identify disordered structures that provide a free spectral range smaller than the frequency shift we can achieve with a single lasing mode (typically 1 nm). Ideally, we need 10-15 modes within 8 to 10 nm spectral bandwidth of the gain curve. Interestingly, by varying the degree of disorder, the spectral density of lasing modes varies, as shown in Fig. 4c where the average number of lasing modes is plotted as a function of disorder deviation from the mean periodic position. Here, a 1000 $\mu$m sample with 125 air grooves has been considered, with increasing disorder ranging from 0.1 $\mu$m to 3 $\mu$m deviation from the mean spatial period of 8 $\mu$m.  The reason for it is that the spatial confinement of the eigenmodes increases with increasing disorder, resulting in an increasing  number of lasing modes able to reach threshold. We found that a disorder pattern with a deviation of $\pm$ 0.5 $\mu$m is enough to yield an average of 13 lasing modes within the wavelength range of  600 nm to 608 nm. We choose a pump length of 450 $\mu$m and run iterative optimization of the pump profile to select 10 lasing modes distributed over the range of 600-608 nm. The thermal-induced spectral shift is recorded for all the modes. Single-wavelength continuous broadband tunability by mode hopping is achieved over 8 nm, as shown in Fig.4(d).

\section{Conclusion}
In this paper, we have used a stable solid-state dye-based random laser to demonstrate temperature-induced tunability in the visible. By enforcing  singlemode operation using pump shaping method, we have shown how temperature-induced change of the refractive index can achieve spectral tunability. In contrast to other mechanisms reported in the literature, our proposed method ensures singlemode tuning, it is non-invasive and does not require any modification of the sample. This random laser offers the freedom to control the free spectral range (FSR) by changing the degree of disorder. We have demonstrated that the emission wavelength of any individually selected lasing mode can be continuously blue-shifted by up to 1 nm when increasing the temperature of the PMMA-DCM layer. By hoping from one mode to another, we have demonstrated remarkable broadband tunability-range of 8 nm. This tunability-range can in principle be further increased by exploring different disorder configurations and dyes.

Such a tunable random laser offers the advantage of simplicity of fabrication, large tunable bandwidth, compact in size, and the ability to operate in harsh environments compared to conventional tunable lasers that might be bulky, or require precise tuning and control mechanisms, and have limited spectral bandwidth.
This single-wavelength tunable random laser holds promising potential for future applications for on-chip tunable laser sources as well as wearable temperature sensors.

\section{Acknowledgement}
We extend our heartfelt appreciation to Dr. Leonid Wolfson for his unwavering dedication to the lab, and to Dr. Yossi Abulafia for his valuable assistance with the fabrication process. We are grateful to the Bar-Ilan Institute of Nanotechnology and Advanced Materials for providing us access to their fabrication facilities.


\section{Disclosures}
The authors declare no conflicts of interest.

\section{Data Availability Statement}
Data underlying the results presented in this paper are not publicly available at this time but may be obtained from the authors upon reasonable request.

{}
\end{document}